\documentclass{aa}
\usepackage{graphicx}

\begin{document}
\renewcommand{\theequation}{\thesection.\arabic{equation}}
\begin{titlepage}

\titlerunning{Microlensing search towards M31}
\title{Microlensing search towards M31}
\authorrunning{S. Calchi Novati et al.}
\author{S. Calchi Novati$^{1}$\footnote{Research supported by fund ex 60 \%
D.P.R. 382/80 and F.S.E. of European Community}, G. Iovane$^{1}$,
A.A. Marino$^{2}$, M. Auri{\`e}re$^{3}$, P. Baillon$^{4}$, A.
Bouquet$ ^{5}$, V. Bozza$^{1}$, M. Capaccioli$^{2}$, S.
Capozziello$^{1}$, V. Cardone$^{1}$, G. Covone$^{6}$, F. De
Paolis$^{7}$, R. de Ritis$^{6}$\footnote{Deceased September 2000},
Y. Giraud-H{\'e}raud$^{5}$, A. Gould$^{5}$\footnote{On leave of
absence from Dept. of Astronomy Ohio State University, Columbus,
OH43210 USA}, G. Ingrosso$^{7}$, Ph. Jetzer$^{8,9}$, J. Kaplan$
^{5}$, G. Lambiase$^{1}$, Y. Le Du$^{5}$, L. Mancini$^{1,8}$, E.
Piedipalumbo$^{6}$, V. Re$^{1}$, M. Roncadelli$^{10}$, C.
Rubano$^{6}$, G. Scarpetta$^{1}$, P. Scudellaro$^{6}$, M.
Sereno$^{6}$, and F. Strafella$^{7}$}
\offprints{novati@sa.infn.it} \institute {$^{1}$ Dipartimento
diFisica ``E.R. Caianiello'', Universit{\`a} degli Studi di
Salerno, and INFN Sez. di Napoli - Gruppo Collegato di Salerno, Italy,\\
$^{2}$ Osservatorio Astronomico di Capodimonte, Napoli, and INFN, Sez. di Napoli, Italy,\\
$^{3}$ Observatoire Midi-Pyren{\'e}es, France,\\
$^{4}$ CERN, 1211 Gen{\`e}ve 23, Switzerland,\\
$^{5}$ Physique Corpusculaire et Cosmologie, Coll{\`e}ge de
France, Paris, France,\\
$^{6}$ Dipartimento di Scienze Fisiche, Universit{\`a} degli
Studi di Napoli ``Federico II'' and INFN, Sez. di Napoli, Italy,\\
$^{7}$ Dipartimento di Fisica, Universit{\`a} di Lecce, Italy,\\
$^{8}$ Institute of Theoretical Physics,
University of Z{\"u}rich, Switzerland,\\
$^{9}$ Institute of Theoretical Physics, ETH, Z{\"u}rich, Switzerland,\\
$^{10}$ INFN Sez. di Pavia, Pavia, Italy.}

\date{Received/ Accepted}
\maketitle

\begin{abstract}
We present the first results of the analysis of data collected
during the 1998-99 observational campaign at the 1.3 meter
McGraw-Hill Telescope, towards the Andromeda galaxy (M31), aimed
to the detection of gravitational microlensing effects as a probe
of the presence of dark matter in our and in M31 halo. The
analysis is performed using the \emph{pixel lensing} technique,
which consists in the study of flux variations of unresolved
sources and has been proposed and implemented by the AGAPE
collaboration. We carry out a shape analysis by demanding that
the detected flux variations be achromatic and compatible with a
Paczy\'nski light curve. We apply the Durbin-Watson hypothesis
test to the residuals. Furthermore, we consider the background of
variables sources. Finally five candidate microlensing events
emerge from our selection. Comparing with the predictions of a
Monte Carlo simulation, assuming a standard spherical model for
the M31 and Galactic haloes, and typical values for the MACHO
mass, we find that our events are only marginally consistent with
the distribution of observable parameters predicted by the
simulation.


\end{abstract}
 \vfill
\end{titlepage}

\section{Introduction} \label{intro}
In the last decade much attention has been focused on the
possibility that a sizable fraction of galactic dark matter
consist of  MACHOs (Massive Astrophysical Compact Halo Object).
Since 1992, the MACHO (Alcock et al. 1993) and EROS (Aubourg et
al. 1993) collaborations have looked towards the Large and Small
Magellanic Clouds (LMC and SMC) in order to detect MACHOs using
gravitational microlensing. This technique, originally proposed
by Paczy\'nski (1986), analyses the luminosity variation of
resolved  source stars, due to the passage of MACHOs close to the
line of sight between the source and the observer.

The MACHO collaboration (Alcock et al. 2000) discovered 13 - 17
microlensing events towards the LMC. Assuming that all events are
due to MACHOs in the halo, about $20\% $ of the halo dark matter
resides in form of compact objects with a mass in the range $0.15$
--- $0.9 M_{\odot}$. The EROS collaboration (Lasserre et al.
2000) observed 6 microlensing events, 5 in the direction of the
LMC and 1 in the direction of the SMC. These observations place an
upper limit on the halo dark matter fraction in the form of
MACHOs. In particular, they exclude, at the $95\%$ confidence
level, that more than $40\%$ of a standard halo is composed of
objects in the range $10^{-7}M_{\odot}$ --- $1 M_{\odot}$. Note
that the results of the two collaborations are consistent with a
$20\%$ halo dark matter fraction  of objects $\sim 0.4 M_{\odot}$.

The OGLE collaboration (Udalski et al. 1993) originally searched
for microlensing events only towards the Galactic bulge, but has
now also extended its search to the LMC and SMC.

A natural extension of the microlensing observational technique
consists in observing dense stellar fields even if single stars
cannot be resolved, as in the case of the M31 galaxy. For this
purpose, the pixel lensing technique has been proposed (Baillon
et al. 1993) and then implemented by the AGAPE collaboration
(Ansari et al. 1997).  Another technique, based on image
subtraction, has been developed by the VATT-Columbia
collaboration (Crotts 1992; Tomaney \& Crotts 1994), and is used
also in the WeCAPP project (Riffeser et al. 2001). The monitoring
of M31 has the advantage that the Galactic halo can be probed
along a line of sight different from those towards the LMC   and
SMC. Furthermore, the observation of an external galaxy  allows
one to study  its halo globally, which, in the case of M31, has a
particular signature due to the tilted disk. Accordingly, the
expected optical depth for microlensing varies from the near to
the far side of the M31 disk (Crotts 1992; Jetzer 1994).

The efficiency of the pixel lensing method to detect luminosity
variations has been tested  by the AGAPE collaboration on data
taken at the 2 meter Bernard Lyot Telescope  in two bandpasses
($B$ and $R$), covering 6 fields of $4.5'\times4.5'$ each around
the center of M31, during 3 years of observations (1994-1996). A
possible microlensing candidate has been observed and further
characterized by using information from an archival Hubble Space
Telescope WFPC2  image (Ansari et al. 1999). An important
conclusion of this analysis is that it is crucial to collect data
in two bandpasses over a long duration with  regular sampling.
Very recently, the POINT--AGAPE collaboration (Auri{\`e}re et al.
2001) announced the discovery of a short timescale candidate
event towards M31. Additional microlensing candidates towards the
same target have been reported by the VATT-Columbia collaboration
(Crotts et al. 2000).

In this paper, we present  results  for the 1998-1999 campaign of
observations at the 1.3 meter McGraw-Hill Telescope, MDM
Observatory, Kitt Peak, towards the Andromeda Galaxy. In $\S$
\ref{pixellensing}, we briefly outline the pixel lensing
technique. $\S$ \ref{setup} is devoted to the description of the
observational campaign and the experimental setup. In $\S$
\ref{reduction} we discuss the data reduction procedure (Calchi
Novati 2000) in some detail, in particular the approach used to
eliminate instabilities caused by the seeing and to evaluate the
errors. In $\S$ \ref{anaml} we present our selection pipeline
(Calchi Novati 2000): bump detection ($\S$ \ref{anabump}), shape
analysis ($\S$ \ref{anashape}) and color and timescale selection
($\S$ \ref{anamira}). We select a sample of 5 light curves that we
retain as microlensing candidate events and whose characteristics
are given in $\S$ \ref{anaevts}. In $\S$ \ref{novae} we show the
light curve of a nova located inside our field of observation:
the discovery of variable sources is a natural byproduct of the
microlensing search. In $\S$ \ref{mcs} we conclude with a
comparison of the outcome of our selection with the prediction of
a Monte Carlo simulation.

\section{Pixel Lensing} \label{pixellensing}

\begin{figure*}
{\includegraphics{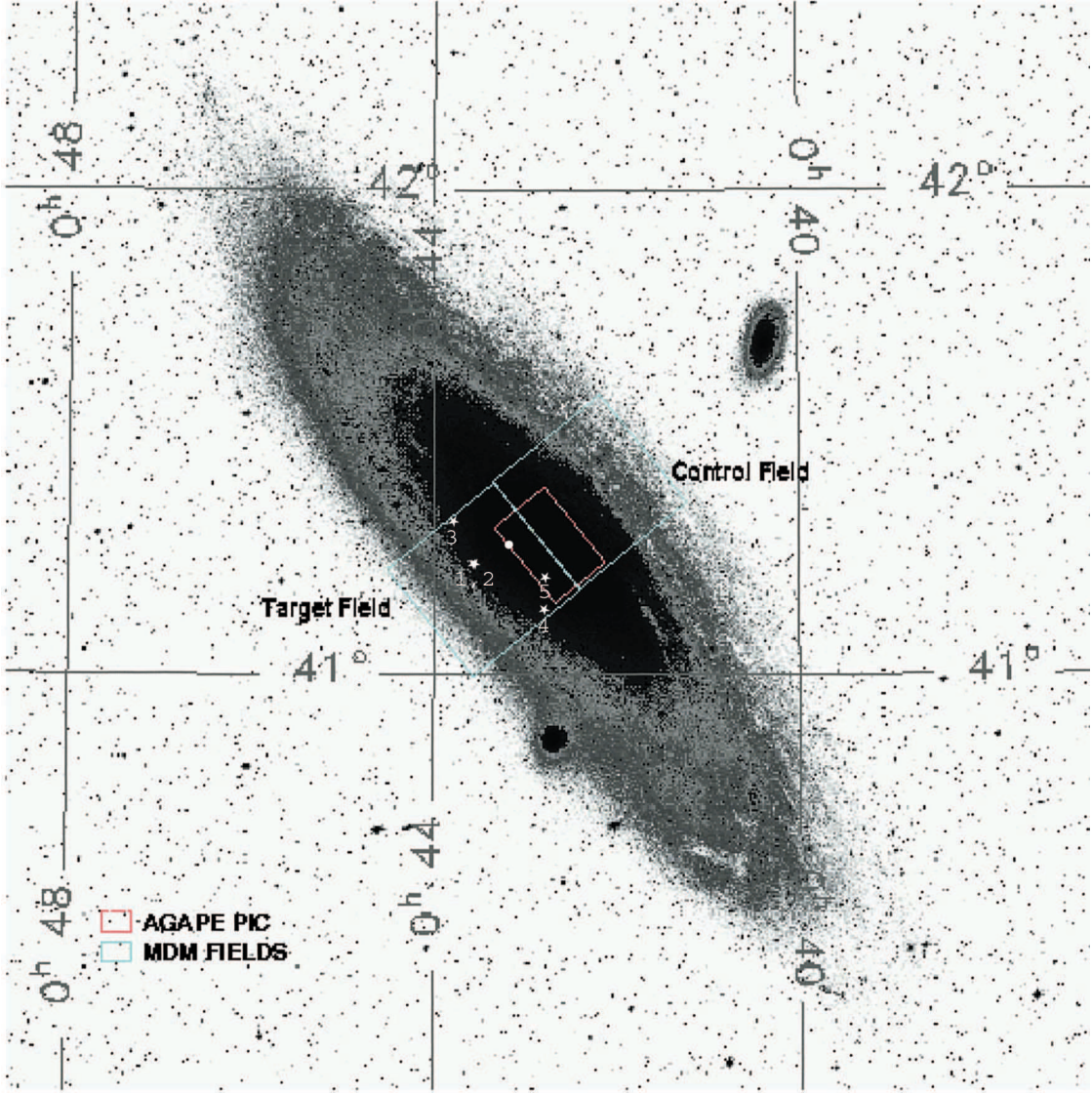}}
{\caption{M31 with MDM and Agape observation fields (courtesy of
A. Crotts). White stars and dots give, respectively, the position
of the five microlensing candidate events (labelled as in Table \ref{tavola2}
and where candidates 1 and 2
appear to be superimposed) and of the nova.}
\label{fig:plotm31}}
\end{figure*}

Pixel lensing is an efficient tool for searching  microlensing
events when the sources can not be resolved. In this case, the
light collected by each pixel is emitted  by a huge number of
stars. Although, in principle, \emph{all} stars in the pixel field
are possible sources, one can only detect lensing events due
either to bright enough stars or to high amplification events.
Typical sources are  red giant stars with $M_I\in[-3.5,0]$. We
estimate that there are about $100$ sources per square arc second
that fit these requirements. The main drawback of the method is
that usually we have no direct knowledge of the flux of the
unamplified source.

For images obtained by the observation of dense stellar fields,
the flux collected by a pixel is the sum of the fluxes emitted by
single stars, which all contribute to the background. If one of
these stars is lensed, its flux varies accordingly. Whenever this
variation is large enough, it will be distinguishable from the
background produced by the other stars. Denoting by $\phi$ the
amplified flux detected by the pixel and by $\phi_{bkg}$ the
background flux, the flux variation is given by
\begin{equation}
 \phi-\phi_{bkg}=\phi^{*}\left[A(t)-1\right]\,,
\end{equation}
where $A(t)$ represents the amplification as a function of time,
\begin{equation}
\phi_{bkg}=\phi^{*}+\phi_{res}\,,
\end{equation}
$\phi^{*}$ is the flux of the star before lensing, and
$\phi_{res}$ is that given by the other stars. In the point-like
and uniform--motion approximations, $A$ is related to the lensing
parameters by:
\begin{equation} \label{amplificazione}
A = \frac{u^{2}+2}{u \sqrt{u^{2}+4}} \, ,
\end{equation}
\begin{equation}
u^{2} = \frac{(t-t_{0})^{2}}{t_{\textrm E}^2} + u^{2}_{0}\,,
\end{equation}
where $t_{{\textrm E}}$  is the  \emph{Einstein time}, $u_{0}$ the
impact parameter in units of  the \emph{Einstein radius}
$R_{{\textrm E}}$, and $t_{0}$ the time of the maximum
amplification. The Einstein radius is
\begin{equation}
R_{{\textrm E}} = \sqrt{\frac{4GM}{c^{2}} \frac{D_{ol}D_{ls}}{D_{os}}}\,,
\end{equation}
where $M$ is the mass of the lens,  $D_{ol}$, $D_{os}$ and
$D_{ls}$ are the observer-lens, observer-source, and lens-source
distances, respectively.

Two characteristic features of a microlensing event are
achromaticity and the uniqueness of its luminosity bump, although
differential amplification of extended sources can give rise to a
chromatic, but still symmetric, lensing light curve (Han et al.
2000).

\section{Observations and experimental setup} \label{setup}

The data analysed in this paper  have been collected on the 1.3
meter McGraw-Hill Telescope, at the MDM observatory, Kitt Peak
(USA)\footnote{Data shared with the Columbia-Vatt
collaboration.}. Two fields have been observed, which lay on the
two sides of the galactic bulge (see Fig. \ref{fig:plotm31}) and
have been chosen in order to be able to study the expected
gradient in the optical depth. The two fields are almost parallel
to the major axis of M31; their centers are located at $\alpha=$
00h43m24s, $\delta = 41^{\!\circ}\,12'\,10''$ (J2000) (named
``Target''), and $\alpha=$ 00h42m14s, $\delta
=41^{\!\circ}\,24'\,20''$ (J2000) (named ``Control''). The data
acquired in the Target field are analysed here.

Fig. \ref{fig:plotm31} shows the  location of the fields and for
comparison also the smaller AGAPE field. The observations were
taken with a CCD camera of $2048 \times 2048$ pixels with
$0.\hskip-2pt ''5$ and therefore a total field size of $17'\times
17'$.

In order to test for achromaticity, images have been taken in two
bands, a wide $R$ and a near--standard $I$. The exposure time is 6
minutes for $R$,  5 minutes for $I$. The observations started in
the fall of 1998 and are still underway.

\begin{figure}
\resizebox{7 cm}{!}
{\includegraphics{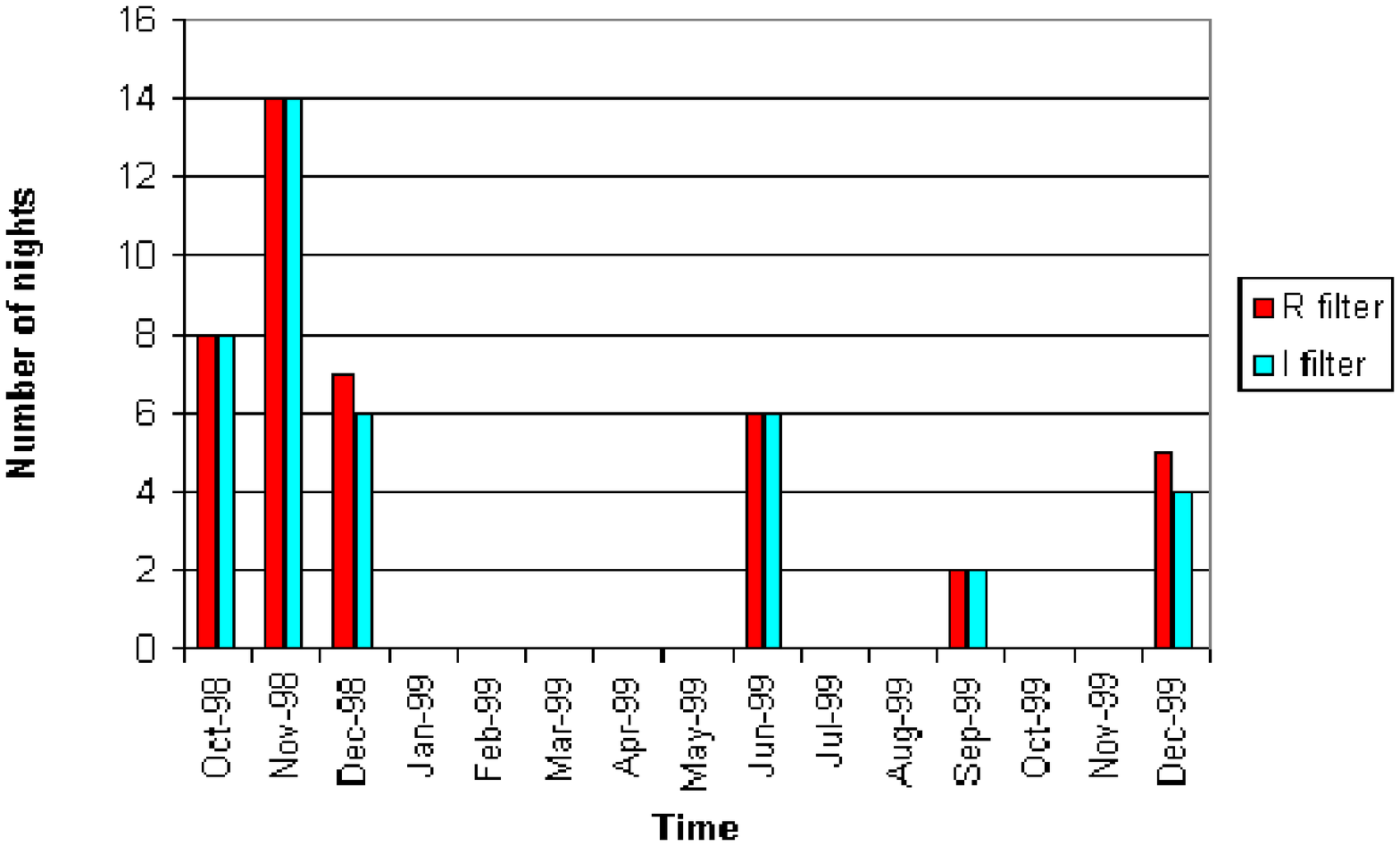}}
\resizebox{7 cm}{!}
{\includegraphics{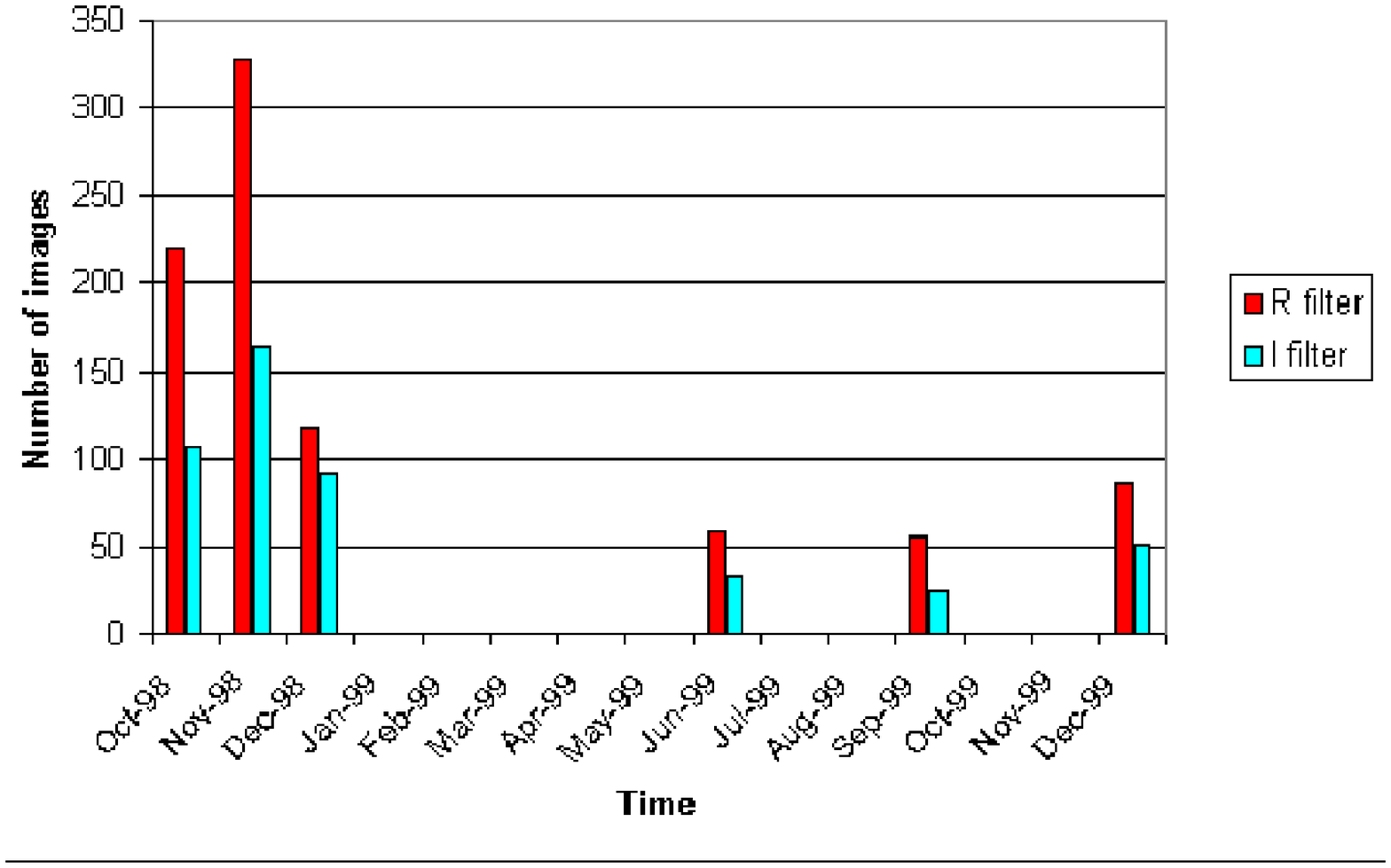}}
\parbox[b]{7 cm}
\vfill
{\caption{Time sampling for the observations in the ``Target''
field.} \label{fig:images}}
\end{figure}

Here we analyse the data taken in the period from the beginning of
October 1998 to the end of December 1999. In Fig. \ref{fig:images}
we give the time sampling of the measurements (number of nights
and images).  For each night we have measurements in both $R$ and
$I$. On average, there are twice as many $R$ images as $I$ images.
In $R$ band we have $\sim 800$ images distributed along 42 nights
of observation.

Most of the observations (about $80\%$) are concentrated in the
first three months, so that, unfortunately, the time distribution
of the data is not optimized for the study of microlensing
effects. Thus, the given time distribution allows us to select
events that take place almost exclusively during the first three
months of observation.  Furthermore, the time coverage of about 14
months is still not long enough to test conclusively the bump
uniqueness requirement for a microlensing event. Mainly for this
reason, we will speak in this paper only of \emph{candidate}
microlensing events.

Taking into account the transmission efficiency of the  filters
and the catalogued magnitudes $R_c$ and $I_c$ (Cousins colour
system), for a sample of $23$ reference secondaries identified in
the Target field (Magnier et al. 1993), we derive the following
photometric calibrations
\begin{equation}\label{calibrazioneR}
R_{c}= m_R-0.13\cdot \left(m_R-m_I\right)+22.54\,,
\end{equation}
\begin{equation}\label{calibrazioneI}
I_{c}=m_I-0.02\cdot \left(m_R-m_I\right)+22.21\,,
\end{equation}
where $m_{R(I)}=-2.5\cdot\log(\phi^*_{R(I)})$ and $\phi^*_{R(I)}$
is the flux of the source in $ADU/s$ measured in the $R$ and $I$
filters respectively. The estimated error is $\simeq 0.1$ mag,
both for $R$ and $I$.

\section{Data  reduction} \label{reduction}

During each night of observation about 20 images are taken in $R$
and 12 in $I$. In principle, this allows two possible strategies
for the analysis. We can study flux variations on light curves
built either by a point obtained from each image, or by a point
obtained by averaging over many images. In the first case, we are
potentially sensitive to very short time variations. However,
this sensitivity is undermined by the low signal-to-noise ratio
(S/N). In the second case, the S/N is increased by the square root
of the number of images we combine.

Results of the analysis of light curves built with one point per
image will be discussed in a future paper. Here we concentrate on
the analysis of light curves obtained after combining all the
images taken in the same night, using a simple averaging procedure
performed on geometrically aligned images (see below).

Data reduction is carried out as follows. After the usual
corrections for instrumental effects, debiasing and flatfielding,
we normalize all the images to a common reference to cope with
variations induced by the observational conditions which are
different from image to image (so that we get \emph{global}
stability conditions on each image with respect to a given one).
We can distinguish three separate effects: the geometric offset
of each image with respect to the others, the difference in
photometric conditions of the sky and seeing effects.

By means of geometrical alignment we obtain that each pixel, on
all the images, is directed towards  the same portion of M31. We
take advantage of the fact that the mean seeing disk is much
larger than the pixel size. We follow Ansari et al. (1997), and
get a precision better than $0.\hskip-2pt ''1$.

Following the methods developed by the AGAPE collaboration
(Ansari et al. 1997),  we then bring all the images to the same
photometric conditions in such a way that the images are globally
normalized to a common reference. The procedure is based on the
hypothesis that a linear relation exists between ``true'' and
measured flux. It then follows that
\begin{equation}\label{fotometria}
  \phi^{pixel}_{ref}(i,j)=a_{curr}\,\cdot\phi^{pixel}_{curr}(i,j) + b_{curr}\,,
\end{equation}
where $\phi^{pixel}_{ref}(i,j)$ and $\phi^{pixel}_{curr}(i,j)$ are the fluxes in
pixel $(i,j)$ for the reference and the current image, and
$a_{curr}$ and $b_{curr}$ are two correction coefficients, which
are the same for all the pixels of the image, that take into
account the effects of variable atmospheric absorption and
variable sky respectively.

The seeing effect gives rise to a spread of the received signal.
In our data the seeing varies from $\sim 1.\hskip-2pt ''3$ up to
$\sim 2.\hskip-2pt ''2$. Consequently we observe fake fluctuations
on light curves obtained after photometrical alignment. In order
to cope with this effect, and thus to get reasonably stable light
curves, we follow a two steps procedure (for further details see
Ansari et al. (1997) and Le Du (2000)). We begin by substituting
the flux of each pixel with the flux of the corresponding
\emph{superpixel}, defined as the flux received on a square of
$m\times m$ pixels around the central one. The value $m$ should
be large enough to cover the typical seeing disk, but not too
large to avoid an excessive dilution of the signal. Given the
mean seeing value and the angular size of the pixel, we choose
$m=5$. This corresponds to $2.\hskip-2pt ''5$, compared to the
average value for the seeing of $\sim 1.\hskip-2pt ''7$ for both
$R$ and $I$ images. In this way we get a substantial gain in
stability since elementary pixels are strongly affected by seeing
fluctuations.

Denoting by $\Phi(i,j)$ the flux in a superpixel, we have
     \begin{equation}
          \Phi(i,j) = \sum_{k = i-n}^{i+n} \sum_{l = j-n}^{j+n}
          \phi^{pixel}(k,l),
     \end{equation}
where $n = (m-1)/2$ and $m$ is the superpixel size.

\begin{figure}
\resizebox{7 cm}{!}
{\includegraphics{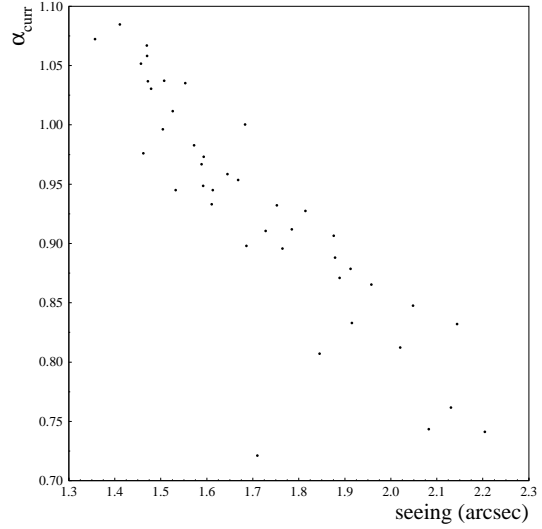}}
\parbox[b]{7 cm} \vfill
{ \caption{The value of the correction factor $\alpha_{curr}$ in
the relation (\ref{seeingcorrection}) for each composed image as
a function of the seeing.} \label{fig:alpha} }
\end{figure}

Instead of  trying to evaluate the point spread function of the
image, as a second step we  apply an empirical stabilization of
the difference between the flux measured on the image and that of
the \emph{median} image, obtained by removing small scale
variations with a median filter on a very large window of
$31\times 31$ pixels (in this way we get an image whose signal is
independent from the seeing value). The stabilization is then
based on the observed linear correlation, for each superpixel,
between these differences measured on the current image (after
photometrical alignment) and the reference image. Denoting with
$\Phi_{ref}(i,j),\,\Phi_{curr,a}(i,j)$ and $\Phi^{med}(i,j)$ the
value of the flux in a superpixel $(i,j)$ for the reference, the
current (photometrically aligned) and the median images
respectively, we have the empirical relation
\begin{eqnarray}\label{seeingcorrection}
{\Phi}_{curr,a}(i,j)-{\Phi}^{med}_{curr,a}(i,j)=\nonumber\\
\alpha_{curr}\,\cdot\left[\Phi_{ref}(i,j)-\Phi^{med}_{ref}(i,j)\right]\,.
\end{eqnarray}
The slope $\alpha_{curr}$, calculated with a minimization
procedure, shows a clear correlation with the seeing (Fig.
\ref{fig:alpha}). This variation is expected because the flux of a
star that enters a superpixel changes with the seeing.

\begin{figure}
\resizebox{7 cm}{!}
{\includegraphics{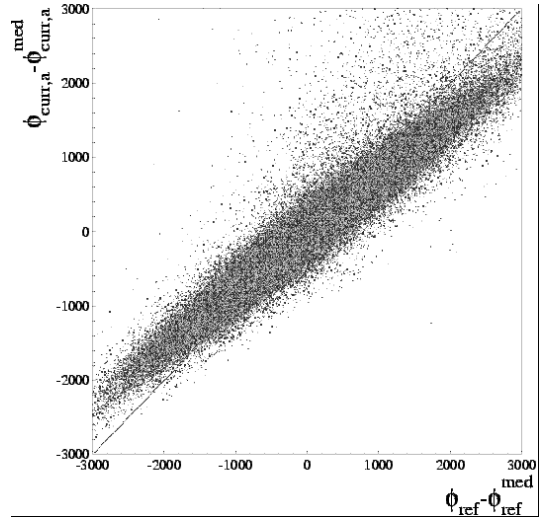}}
\parbox[b]{7 cm} \vfill
{\caption{Plot showing the linear correlation between the
quantities
$\left[\Phi_{curr,a}(i,j)-\Phi_{curr,a}^{med}(i,j)\right]$ and
$\left[\Phi_{ref}(i,j)-\Phi_{ref}^{med}(i,j)\right]$. The dashed
line is the $y=x$ line.} \label{fig:cigar}}
\end{figure}

In the example shown in Fig. \ref{fig:cigar} the seeing of the
current image is greater than that of the reference image, and we
find $\alpha_{curr}<1$. We note that in this case, the flux of
superpixels for which $\Phi_{curr,a}<\Phi_{curr,a}^{med}$ are
corrected to a lower value (points in the bottom left corner),
while, if $\Phi_{curr,a}>\Phi_{curr,a}^{med}$, they are corrected
up (points in the up right corner). If the seeing of the current
image is smaller than that of the reference image
($\alpha_{curr}>1$) the situation is reversed.

\begin{figure}
\resizebox{7 cm}{!}
{\includegraphics{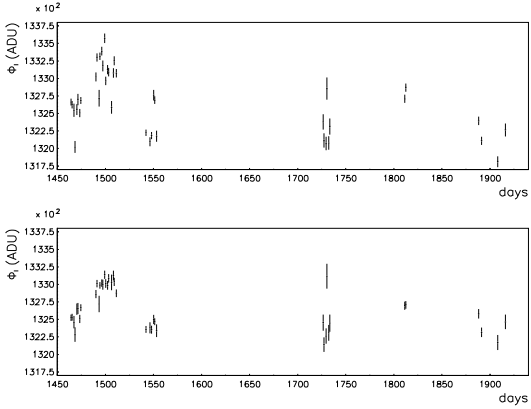}}
\parbox[b]{7 cm} \vfill
{\caption{The same light curve before (top) and after seeing
correction. In both cases the error bar shows just the photon
noise.} \label{fig:seeingcl}}
\end{figure}

While stressing its empirical character, we note that this
approach is rapid and efficient. In Fig. \ref{fig:seeingcl} we
show the effect of the correction on a given light curve.

To construct a corrected current pixel flux as close as possible
to the reference flux, $\Phi_{ref}$, we replace $\Phi_{ref}$ in
(\ref{seeingcorrection}) by the corrected current flux
$\hat{\Phi}_{curr}$ and solve for the latter: 
\begin{eqnarray}\label{seeingcorrected}
  \hat{\Phi}_{curr}(i,j)-\Phi^{med}_{ref}(i,j) =\nonumber \\
  \frac{1}{\alpha_{curr}} \,\cdot
  \left[\Phi_{curr,a}(i,j) -\Phi^{med}_{curr,a}(i,j)\right]\,.
\end{eqnarray}
That is, the corrected flux for the current image is given by the
sum of the median of the reference image, and of the weighted
deviation from its median, i. e. by its characteristic small
spatial scale variations, depending on the relative absorption and
on the seeing conditions\footnote{We have verified that it is
possible to exploit the relation (\ref{seeingcorrection}) with
${\Phi}_{curr,a}$ replaced by $\Phi_{curr}$ (i.e. using non
photometrically aligned images) in order to perform in one single
step the photometrical alignment and the seeing correction. In
the two cases we get the same final results.}.

\begin{figure}
\resizebox{7 cm}{!}
{\includegraphics{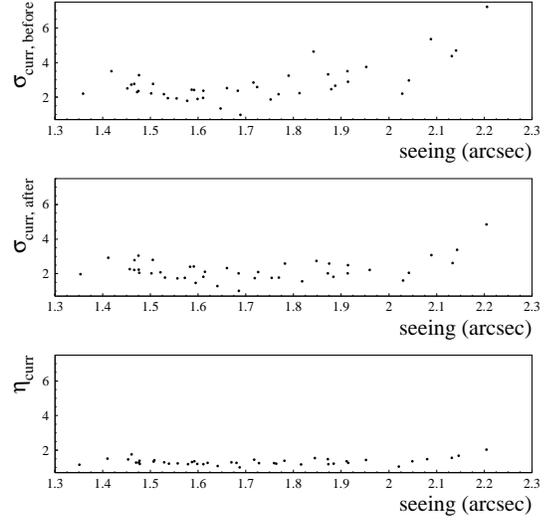}}
\parbox[b]{7 cm} \vfill
{ \caption{The dispersion for the
distribution (\ref{gaussiana}) calculated for each composed image
as a function of the seeing before ($\sigma_{curr,before}$) and
after ($\sigma_{curr,after}$) seeing correction (taking all the
points of the image) and for the sample of points selected
according to the condition (\ref{stability}).} \label{fig:dispe} }
\end{figure}

In order to minimize the deviations from the median, we choose as
a reference image (one for each filter), an image characterized by
a seeing value equal to the average value over the period of
observations. The seeing fraction of a source in a superpixel is
then $\simeq 0.87$.

Another crucial point  is the evaluation of the error to be
associated with the received flux. In order to give a more
appropriate, though empirical, error estimate, we renormalize the
photon noise $\sigma_{stat}$ by introducing a correction factor
$\eta_{curr}$, that depends on the image, to include all the
systematic effects over which we have less control. For a
discussion on the relation between the photon noise and others
different systematic effects such as surface brightness
fluctuations see Gould (1996).

The evaluation of $\eta_{curr}$ is based on the study of the
dispersion of the distribution of the normalized difference,
superpixel by superpixel, between the current and the reference
image
\begin{equation}\label{gaussiana}
\frac{\hat{\Phi}_{curr}(i,j)-\Phi_{ref}(i,j)}
{\sigma_{stat}[\hat{\Phi}_{curr}(i,j)-\Phi_{ref}(i,j)]}\,,
\end{equation}
where the denominator is the statistical error on the difference.

This distribution is expected to have zero mean (which follows
from the geometrical and photometrical alignment) and dispersion
one (which would indicate that the photon noise alone gives the
right evaluation of the error). We do find a null mean but the
dispersion is greater than one and depends on the seeing value. We
note, however, that this effect is greatly reduced by the seeing
correction (see Fig. \ref{fig:dispe}). For the reference image we
assume that the estimated error $\sigma_{est}[\Phi_{ref}]$
coincides with the statistical error given by the photon noise,
\begin{equation}
  \sigma^2_{est}\left[\Phi_{ref}(i,j)\right]=
  \sigma^2_{stat}\left[\Phi_{ref}(i,j)\right]\,,
\end{equation}
because it does not suffer from additional noise due to
geometric, photometric and seeing transformations, as do the
other images. On the other hand, this means that we evaluate the
estimated error with respect to that of the reference image.

The correction factor $\eta_{curr}$ is then equal to the
dispersion of the distribution (\ref{gaussiana}) calculated for a
sample of points, properly selected according to the criterion
that they belong to ``stable'' light curves, in order
to exclude light curves which show real
stellar flux variations. If the subset is
small enough and homogeneous, the correction factor appears to
be, as expected, almost independent of the seeing value, and has
an average value $\eta_{curr} \sim 1.4$ (Fig. \ref{fig:dispe},
bottom). The estimated error of the $(i,j)$ superpixel is
obtained from the relation
\begin{eqnarray} \label{errore}
\sigma^2_{est}\left[\hat{\Phi}_{curr}(i,j)-\Phi_{ref}(i,j)\right]= \nonumber \\
\eta^2_{curr}\,\cdot\sigma^2_{stat}\left[\hat{\Phi}_{curr}(i,j)-\Phi_{ref}(i,j)\right]\,.
\end{eqnarray}

The ``stable'' pixels are selected by imposing the condition
\begin{equation} \label{stability}
  \frac{\bar{\Phi}(i,j)-\bar{\Phi}_{bkg}(i,j)}
  {\sigma_{stat}[\bar{\Phi}(i,j)-\bar{\Phi}_{bkg}(i,j)]} \le
  \epsilon\,,
\end{equation}
where $\bar{\Phi} (i,j)$ is the average flux along the light
curve, $\bar{\Phi}_{bkg}(i,j)$ is the \emph{baseline} level and
$\sigma_{stat}[\bar{\Phi}(i,j)-\bar{\Phi} _{bkg}(i,j)]$ is the
error associated with the evaluation of the averages. The
baseline is defined as the minimum value taken by the average
flux calculated along $q$ consecutive points on the light curve.
We take $q=6$, in order to avoid underestimation of the baseline
due to fluctuations.  Choosing $\epsilon \sim 1.5$, the selected
points are $\sim 3 \%$ of the total.

\section{Candidates selection} \label{anaml}

A microlensing event is characterized by specific features that
distinguish it from other, much more common, types of luminosity
variability, the main background to our search. In particular for
a microlensing event the bump
\begin{itemize}
  \item does not repeat;
  \item follows the (symmetric) Paczy\'nski shape;
  \item is achromatic.
\end{itemize}

Variable stars usually show multiple flux variations and have an
asymmetric chromatic shape. Moreover, different classes of
variable stars are characterized by specific features, such as
timescale variation and color, that can be used to distinguish
them from real microlensing events.

In the following we devise selection techniques that make use of
these characteristics while taking account of the specific
features of our data set.

\subsection{Bump detection} \label{anabump}

As a first step we select light curves showing a single flux
variation. We begin by evaluating a baseline, i.e. the background
flux ($\bar{\Phi}_{bkg}$) along each light curve, as defined in
$\S$ \ref{reduction}.

Once the baseline level has been fixed, we look for a significant
bump on the light curve. This is identified whenever at least 3
consecutive points exceed the baseline  by $3\,\sigma$. The
variation is considered to be over when 2 consecutive points fall
below the $3\,\sigma$ level. Under the hypothesis that the points
follow a gaussian distribution around the baseline, we use the
estimator $L$, the likelihood function, to measure the
statistical significance of a bump. We want to give more weight to
points that are unlikely to be found, so that we define $L$ as
\footnote{In order to simplify the notation, 
hereafter we write $\Phi$ for
the corrected superpixel flux $\hat{\Phi}_{curr}$ as given in equation \ref{seeingcorrected}.}
\begin{equation} \label{likelihood}
L = -\ln\left(\Pi_{j\in bump}P(\Phi|\Phi>\Phi_{j})\right)
\;\;\mbox{given}\;\; \bar{\Phi}_{bkg},\, \sigma_{j}\,,
\end{equation}
where
\begin{equation}
P(\Phi|\Phi>\Phi_{j}) =  \int_{\Phi_{j}}^\infty d\Phi {\frac{1}{
\sigma_{j}\sqrt {2\pi}}} \exp\left[{-\frac{(\Phi
-\bar{\Phi}_{bkg})^2}{2\sigma_{j}^{2}}}\right].
\end{equation}
$L$  is then a growing function of the unlikelihood that a given
variation is the product of random noise. This estimator is
different from the usual definition leading to a $\chi^{2}$ with
$n$ points, and has the advantage to give weight only to the
positive deviations above threshold, which are the ones of
interest.

For each light curve we denote by $L_{1}$ and $L_{2}$ the two
largest deviations, respectively. We fix a threshold $L_{tresh}$,
and we require $L_{1}> L_{tresh}$ to distinguish real variations
from noise. Moreover, we fix an upper limit to the ratio
$L_{2}/L_{1}$ to exclude light curves with more than one
significant variation. The shape analysis is then carried out on
the superpixels that have the highest values of $L$ in their
immediate neighborhood since we find a cluster of pixels
associated with each physical variation. This method suffers from
a possible bias introduced by an underestimation of the baseline
level (which we further analyse in the next section).

We have carried out a complete analysis selecting the pixels with
the following criteria:
\begin{itemize}
  \item exclusion of resolved stars;
  \item $L_{tresh}=100$;
  \item $L_2/L_1<0.1$.
\end{itemize}
This selection is made only on $R$ images in order to reduce
contamination by variable sources. In this way we take advantage
of the fact that most luminous variables (to which we are anyway
sensitive) show stronger variations in the $I$ than $R$ band.

By using these peak detection criteria, the number of superpixels
is reduced from $\sim 4\cdot 10^6$ to $\sim 5 \cdot 10^3$.

\subsection{(Achromatic) shape analysis} \label{anashape}

As a second step we determine whether the selected flux variation
is compatible with a microlensing event.

The light curve of a microlensing event with amplification $A(t)$
due to a  source star with unlensed flux $\phi^{*}$ (now to be
evaluated in a superpixel) is
\begin{equation}\label{lightcurve}
  \Phi (t)=\Phi_{bkg}+\left(A(t)-1\right) \,
  \phi^{*}\, ,
\end{equation}
where $\Phi (t)$ represents the flux collected in the superpixel
associated with a single pixel, as defined before, and $A(t)$ is
given by (\ref{amplificazione}).

Actually, one can not directly and easily measure $\phi^{*}$, the
unamplified flux of the unresolved source star. Only a
combination of the 5 parameters that characterize the light curve
can be measured in a straightforward manner:
\begin{itemize}
\item $\Phi_{bkg}$, the background level (which include the flux of the unamplified source);
\item $t_0$, the time  of maximum amplification;
\item $t_{1/2}=t_{1/2}\left(t_{\textrm E},\,u_{min}\right)$,
    the time width of the bump at  half--maximum;
\item  $\Delta \Phi_{max} = \Delta \Phi_{max}\left(\phi^{*},\,u_{min}\right) =
    \phi^{*}\, (A_{max}-1)$, the excess of the flux with respect to the background at maximum.
\end{itemize}

Whenever the amplification is high enough, one can approximate
$A\left(t\right)\simeq 1/u\left(t\right)$ and $t_{1/2}\simeq
2\sqrt{3}\,t_{\textrm E}\,u_{min}$. It is then possible to rewrite
the expression (\ref{lightcurve}) in terms of these 4 parameters,
and a degeneracy arises among the parameters of the amplification
$u_0$ and $t_{\textrm E}$, and the unknown flux of the unamplified
source, $\phi^*$ (``degenerate'' Paczy\'nski curve Gould (1996)).
Because of this degeneracy it is in general difficult to get,
without extra-information, a reliable insight into all the
parameters that characterize the light curve, the Einstein time
in particular. For this reason we can extract only the  4
aforementioned parameters even though we carry out a non linear
fit with the complete 5 parameters (``non-degenerate'')
Paczy\'nski curve.

We now refine the selection based on the likelihood estimator in
order to remove unwanted light curves with low S/N and for which
the available data do not allow us to well characterize the bump.
To this end we perform a Paczy\'nski 5--parameters  fit and we
study
\begin{itemize}
\item the signal to noise ratio for the $R$ flux variation;
\item the sampling of the data on the bump.
\end{itemize}

We define the S/N estimator as
\begin{equation}\label{rapp}
  Q\equiv \frac{\chi^2_{const}-\chi^2_{ml}}{\chi^2_{ml}/n.d.f}\,,
\end{equation}
where $\chi^2_{const}$ is the $\chi^2$ with respect to a constant
flux and $\chi^2_{ml}$ is the $\chi^2$ with respect to the
Paczy\'nski fit.

The ratio $Q$ is actually correlated with the likelihood
estimator $L$ we used in the previous step. In parallel with  the
cut $L_1>100$ we then keep only light curves with $Q>100$.

We do not ask for the $I$ bump to be significant.

The second point concerns the necessity to well characterize the
bump shape in order to recognize it as a microlensing event in the
presence of highly irregular time sampling of data (see Fig.
\ref{fig:images}). For this purpose we require  at least 4 points
on both sides of the maximum, and at least 2 points inside the
interval $t_0\pm t_{1/2}/2$.

After this selection we are left with 1356 flux variations.

From now on we work with the data in both colors ($R$ and $I$)
and we carry out a shape analysis of the light curve based on a
two steps procedure as follows:
\begin{itemize}
\item $\chi^2$ selection criterion;
\item Durbin-Watson test on residuals;
\end{itemize}
which we now discuss in some detail.

The first point is taken into account by performing the
non-degenerate Paczy\'nski fit in both colors simultaneously, so
that we check also for achromaticity of the selected luminosity
variations. In particular, we require that the three geometrical
parameters that characterize the amplification ($t_{\textrm
E},\,u_0$ and $t_{0}$)  be the same in both colors. We get,
therefore, a 7 parameters least $\chi^2$ non linear fit:
\begin{equation}
 \chi^2=\sum_{j=1}^{2}\sum_{i=1}^{N_{j}}\frac{\left[\Phi_j(t_{i})-\Phi_{model}
(t_{i}|\Phi^j_{bkg},\phi^{*,j}, t_0,t_{\textrm E},u_0)
\right]^2}{\sigma_{i,j}^2}\,.
\end{equation}

To retain a light curve as a candidate microlensing event, we
require that the reduced $\chi^2$
\begin{equation} \label{cutchi2}
\frac{\chi^2}{N-7}\equiv \tilde\chi^{2}< 1.5
\end{equation}
where $N = N_{R}+N_{I}$ is the total number of points in $I$ and
$R$.

The application of the $\chi^2$ criterion test reduces the sample
of light curves from 1356 to 27.

As a further step we apply the Durbin--Watson (Durbin \& Watson
1951) test to the residuals  with respect to the 7-parameters
non-degenerate Paczy\'nski fit.  With the DW test we check the
null hypothesis that the residuals are not timely--correlated by
studying possible correlation effects between each residual and
the next one against type I error (i.e. against the error to
reject the null  hypothesis although it is correct, e.g. Babu \&
Feigelson (1996)). We require a significance level of 10\%. As
\emph{time} plays a fundamental role in the DW test, we perform
this test on each color separately.

We call $dw_{R}$ and $dw_{I}$ the coefficients for the
Durbin--Watson test on the full data set. In order to retain a
light curve we require $1.54<dw_{R(I)} < 2.46$, appropriate for
40--42 points along the light curve. 

This statistical analysis reduces our sample of light curves from
27 to 11.

It is worthwhile
to note that some light curves, showing a real microlensing event
superimposed on a signal due to some nearby variable source,
and passing the previous selection criteria, could
be excluded by the DW test,  sensitive to
timely correlated residuals.

In order to test our efficiency with respect
to the introduction of the DW test we have done
a study on ``flat'' light curves performing a ``constant flux''
fit, and selecting light curves requiring a reduced $\tilde{\chi}^2<1.5$.
By applying the DW test we then reject 
about 50\% of these light curves, i.e., more than the 10\%
we could expect if we had just random fluctuations. In the
discussion of the Monte Carlo simulation we duly
take into account this effect.

\subsection{Color and timescale selection} \label{anamira}

\begin{center}
\begin{table*}
\begin{tabular}{|c|c|c|}
 
\hline
  criterion & $\%$ pixel excluded & pixel left \\ \hline
  exclusion of resolved stars & $\sim 10\%$ & $\sim 3.6\cdot 10^6$ \\
  mono bump likelihood analysis & $\sim 99.8\%$ & 5269 \\
\hline
  signal to noise ratio ($Q>100$) & $\sim 69\%$ & 1650  \\
  sampling of the data on the bump & $\sim 18\%$ & 1356 \\
\hline
  $\chi^2< 1.5$ & $\sim 98\%$ & 27 \\
  $1.54<dw_{R(I)}<2.46$ & $\sim 59\%$ & 11\\
  \hline
  $t_{1/2}<40$ d or $R-I<1.0$ & $\sim 55\%$ & 5 \\ \hline
\end{tabular}
\caption{Summary of selection criteria.} 
\label{tavola1}
\end{table*}

\begin{table*}
\begin{tabular}{|c|c|c|c|c|c|c|c|c|c|c|}
\hline id & $\alpha$(J2000) & $\delta$(J2000) & $t_{1/2}$ (d) &
$t_0$ (J-2449624.5) & $R_{max}$ & $R-I$ &
$\tilde{\chi}^2$ & $dw_R$ & $dw_I$ \\
\hline
1& 00h43m27.4s  & $41^\circ\, 13'\, 11''$ & $32\pm 6$ & $1506\pm 1$ &
$22.6\pm 0.1$ & $0.3\pm 0.2$ & 1.25 & 1.78 & 1.65  \\ 
2& 00h43m26.5s  & $41^\circ\, 13'\, 16''$ & $22\pm 7$ & $1508\pm 1$ &
$22.7\pm 0.1$ & $0.2\pm 0.2$ & 1.37 & 1.57 & 1.65  \\
3& 00h43m39.9s  & $41^\circ\, 18'\, 41''$ & $39\pm 9$ & $1505\pm 1$ &
$22.2\pm 0.1$& $0.8\pm 0.2$ & 1.48 & 1.97 & 1.67 \\ 
4& 00h42m39.3s & $41^\circ\, 6'\, 53''$ & $15\pm 1$ & $1470\pm 1$ &
$22.3\pm 0.1$ & $0.5\pm 0.2$ & 1.42 & 1.68 & 1.95 \\
5& 00h42m39.1s  & $41^\circ\, 11'\, 26''$ & $25\pm 3$ & $1501\pm 1$ &
$21.7\pm 0.1$ &$0.5\pm 0.2$ & 1.16 & 1.82 & 1.99 \\ 
  \hline
\end{tabular}
\caption{Characteristics of microlensing candidates.} \label{tavola2}
\end{table*}
\end{center}

By far, the most efficient way to get rid of multiple flux
variations due to variable stars is to acquire data that are
distributed regularly for a sufficiently long period of time.
Unfortunately, at present, the data cover with regularity only the
first three months of observation, and the total baseline is less
than 2 years long.

For this reason, in addition to the analytical treatment that
looks for the compatibility with an achromatic Paczy\'nski light
curve (efficient, for instance, to reject nova-like events), we
introduce another criterion based on the study of some physical
characteristics of the selected flux variations.

In particular we note that long period red variable stars (such
as Miras) could not be completely excluded by the selection
procedure applied so far. By contrast, short period variable stars
are eliminated thanks to the cut on the second bump of the
likelihood function. A preliminary analysis of the period, the
color and the light curves of long period red variable stars,
taken from de Laverny et al. (1997), lead us, with a rather
conservative approach, to exclude those light curves that present
at the same time a duration $t_{1/2}>40$ days and a color
$(R-I)_{C}>1.0$. A more detailed analysis aimed at a better
estimation of that background noise is currently underway.

We are aware that this last selection criterion could eliminate
some real microlensing events. The probability that this might
happen is however low because the microlensing timescales are
expected to be uncorrelated with the source color. In fact, a
combination of a large $t_{1/2}$ and color $(R-I)_C>1.0$ is
extremely unlikely for microlensing events, but quite common for
red variables.

This last criterion further reduces the number of candidates from
11 down to 5.

\subsection{Results of microlensing search} \label{anaevts}

We now summarize (Table \ref{tavola1}) the different steps in the selection, give the
number of surviving pixels after the application of the indicated
criteria and the details of our set of microlensing candidate
events.

We take these 5 light curves, whose characteristics we are now
going to discuss, as our final selection of microlensing
candidate events.

In the following table (Table \ref{tavola2}) we  give their position, the estimated
$t_{1/2}$ in days,  the time of maximum amplification $t_0$ (J-2449624.5),
the magnitude at maximum\footnote{Evaluated
starting from the excess of flux with respect to the background
$\Delta\Phi_{max}$.} $R_{max}$ and the color $(R-I)_C$. We then
give the values of the fit: the reduced $\tilde{\chi}^2$ and the
values of the Durbin--Watson $dw_{R(I)}$ coefficients.

\begin{figure*}
{\includegraphics{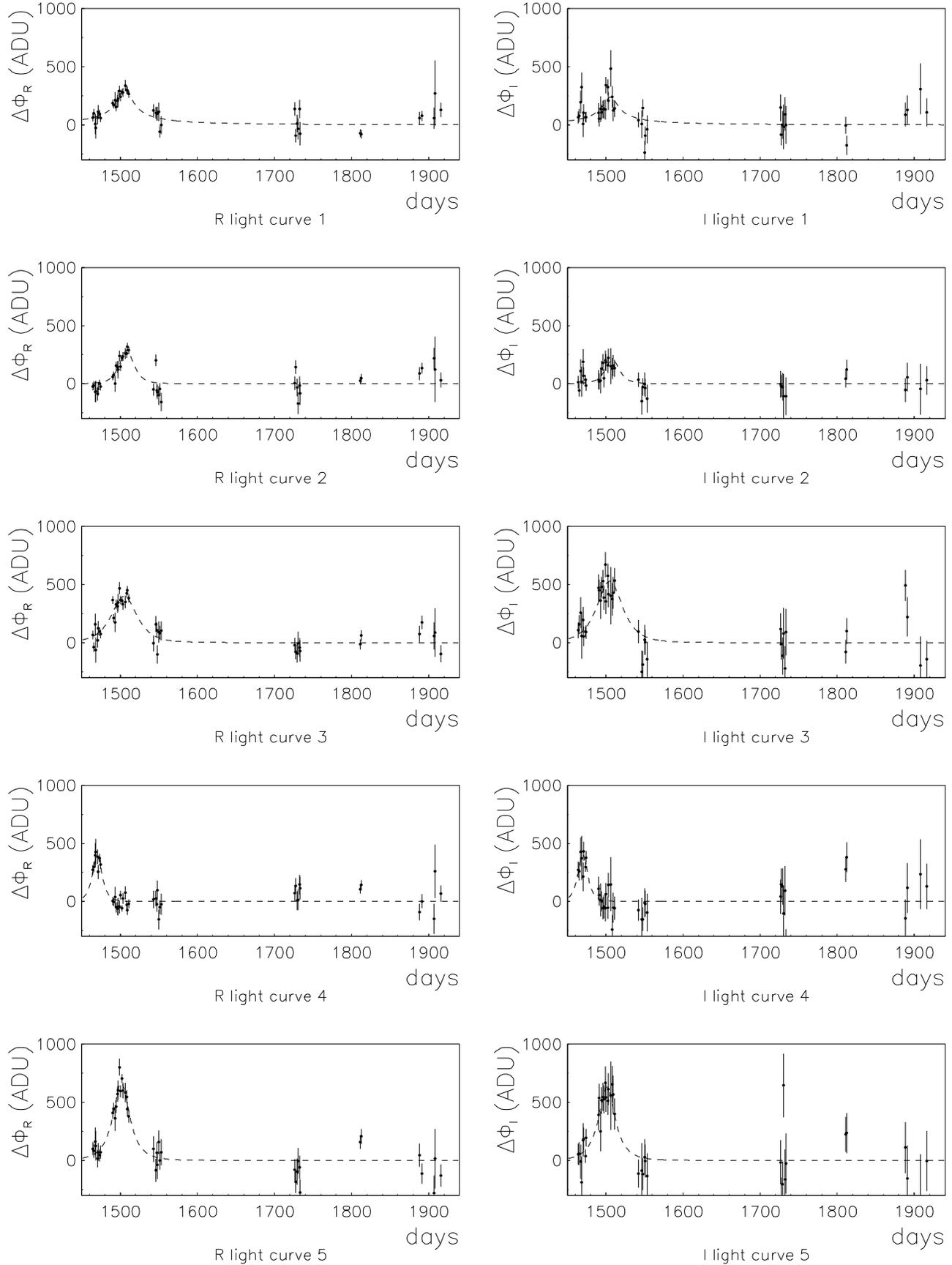}}
{ \caption{Light curves of the 5 candidate microlensing events.
On the y axis, $\Delta\Phi\equiv \Phi-\Phi_{bkg}$. On the x axis,
the origin of time is in J-2449624.5. The dashed line represent
the result of the 7--parameters Paczy\'nski fit.}
\label{fig:5evts} }
\end{figure*}

The corresponding light curves are shown in Fig. \ref{fig:5evts}.

\subsection{Variable stars} \label{novae}

\begin{figure}
\resizebox{7 cm}{!}
{\includegraphics{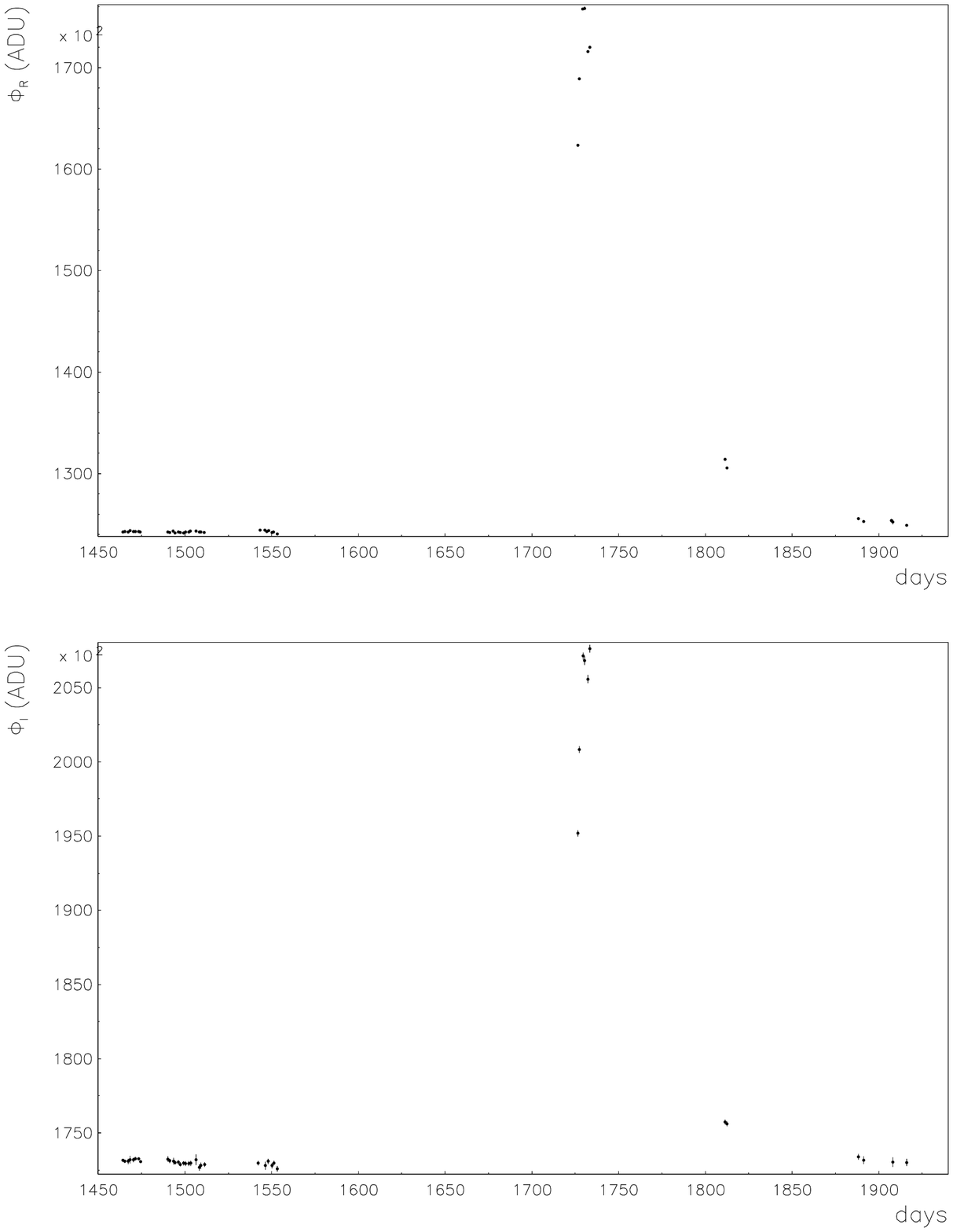}}
\parbox[b]{7 cm} \vfill
{\caption{The $R$ and $I$ light curve of the nova event in
$\alpha$=00h43m1.7s,\,$\delta=41^\circ\, 15'\, 37''$ (J2000).}
\label{fig:novacl}}
\end{figure}

Our data contain many more varying light curves that are due not
to microlensing events but to other variable sources. The study of
these variable stars is an interesting task in itself. Clearly,
pixel lensing is well suited for this research. We give here (see
Fig. \ref{fig:novacl}) only the light curve of an event
characterized by a very strong and chromatic flux variation, that
has already been considered as due to a nova (Modjaz \& Li 1999)
and whose light curve is also shown in Riffeser et al. (2001). We
note that our points, even if with a poorer sampling in the
descent, are in good agreement with those of this collaboration.
This nova is found in the region of the bulge of M31 and we
localize it in $\alpha$=00h43m1.7s,\,$\delta=41^\circ\, 15'\,
37''$ (J2000). We evaluate the magnitude at maximum amplification
as $R_c=17.0$ and $I_c=16.8$.

\section{Discussion of microlensing search} \label{mcs}

In order to gain an insight into the results obtained, we compare
our 5 candidate microlensing events with the prediction of a Monte
Carlo simulation which takes into account  the experimental set
up and the time sampling of the observations. We assume a
standard model (isothermal sphere with a core radius of 5 kpc)
for the haloes of  both M31 and  the Milky Way. The total mass of
M31 is assumed to be twice that of our Galaxy. MACHOs  can be
located in either haloes. Moreover, we consider also self--lensing
due to stars in the M31 bulge or disk. We fix the lens masses at
different values for MACHOs in the halo and stars in the bulge or
disk.  The model of the bulge is taken in Kent (1989), the luminosity function in Han et al. (1998). 
The luminosity function of the disk is determined considering two models:
the one developed in Devriendt et al. (1999) and the model
obtained considering the data of the solar neighborhood taken in Allen (1973) 
corrected for high luminosities (Hodge et al 1988). The results we obtain are almost
insensitive to the particular choice between these two models.

We choose the mass of the lenses in the bulge to be
$m_{MACHO}^{bulge}=0.4\,M_\odot$, and the mass for MACHOs in the
haloes equal either to $m_{MACHO}^{halo}=0.5\,M_\odot$ or to
$m_{MACHO}^{halo}=0.01\,M_\odot$. In both cases about $90\%$ of
the expected events are due to lenses located in M31. Taking into
account our selection criteria, the results for the expected
number of events for a halo fully composed of MACHOs, including a
contribution due to lensing by stars of the bulge and disk of M31
of $\sim 1$ event independent of halo parameters, is $\sim 4$
and $\sim 9$, respectively. The Monte Carlo simulation does not
yet include the effect of secondary bumps due to artifacts of the
image processing (alignment, seeing stabilization) and to
underlying variable objects. From the data, we estimate that
these effects reduce the number of observed events by at most
30\%, and this for the shortest events.

We expect, and this is confirmed by simulations, that the
sources of most detectable events are red giants and 
have very large radii. Therefore, finite size and
limb darkening effects are important, in particular
for low mass lenses. These effects are included in
the simulations\footnote{The finite size effect is computed
exactly from the formul{\ae} by Witt \& Mao (1994). For the limb
darkening we use an approximation method based on Gould (1994).}.
However, in both the real and simulated analysis, we do not
include finite size and limb darkening effects in the 
amplification light curve fitted to the events. This results
in a loss of detection efficiency smaller than 5\%, which
is taken into account in the simulations.

\begin{figure*}
\resizebox{7 cm}{!}
{\includegraphics{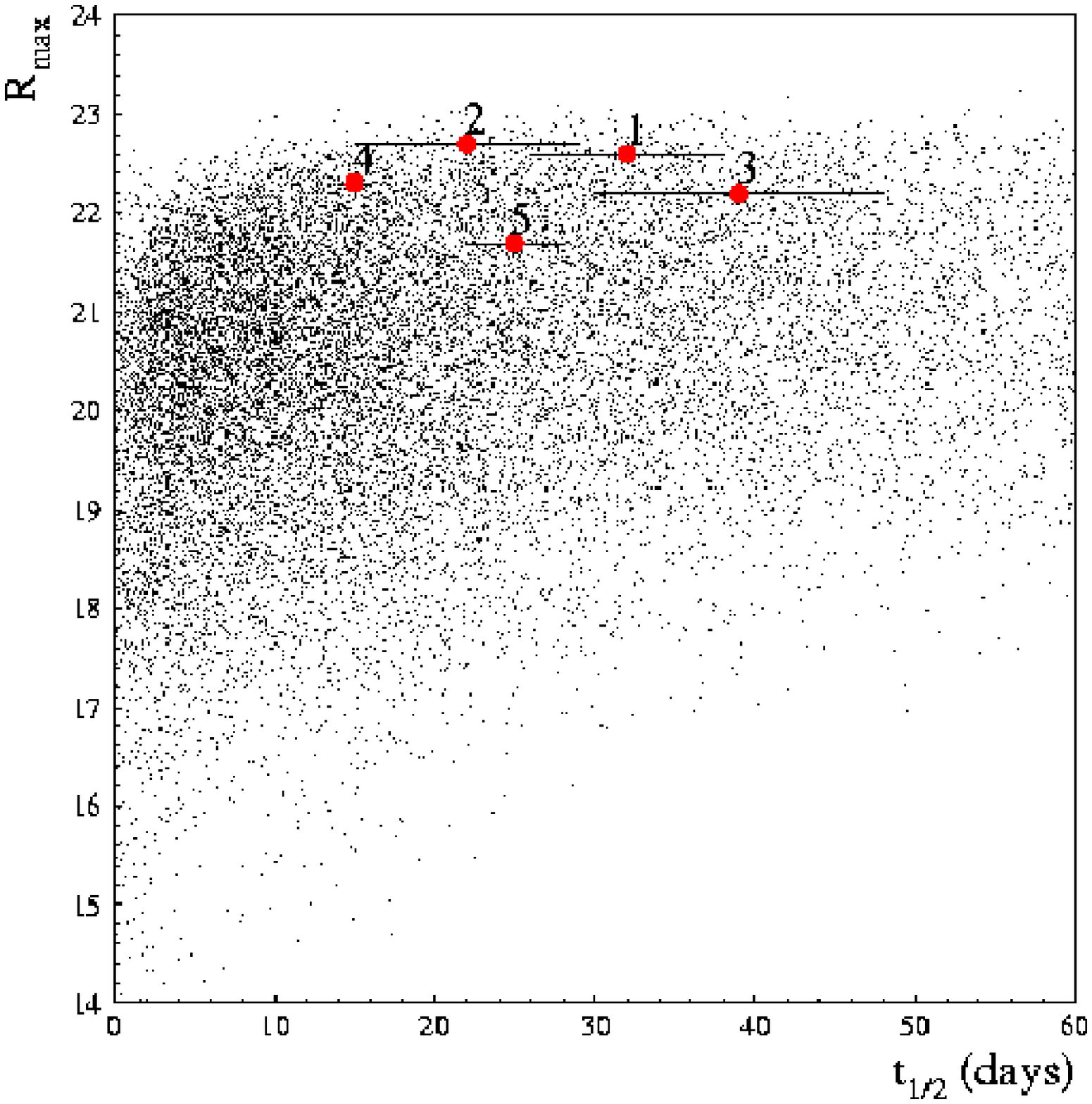}} 
\resizebox{7 cm}{!}
{\includegraphics{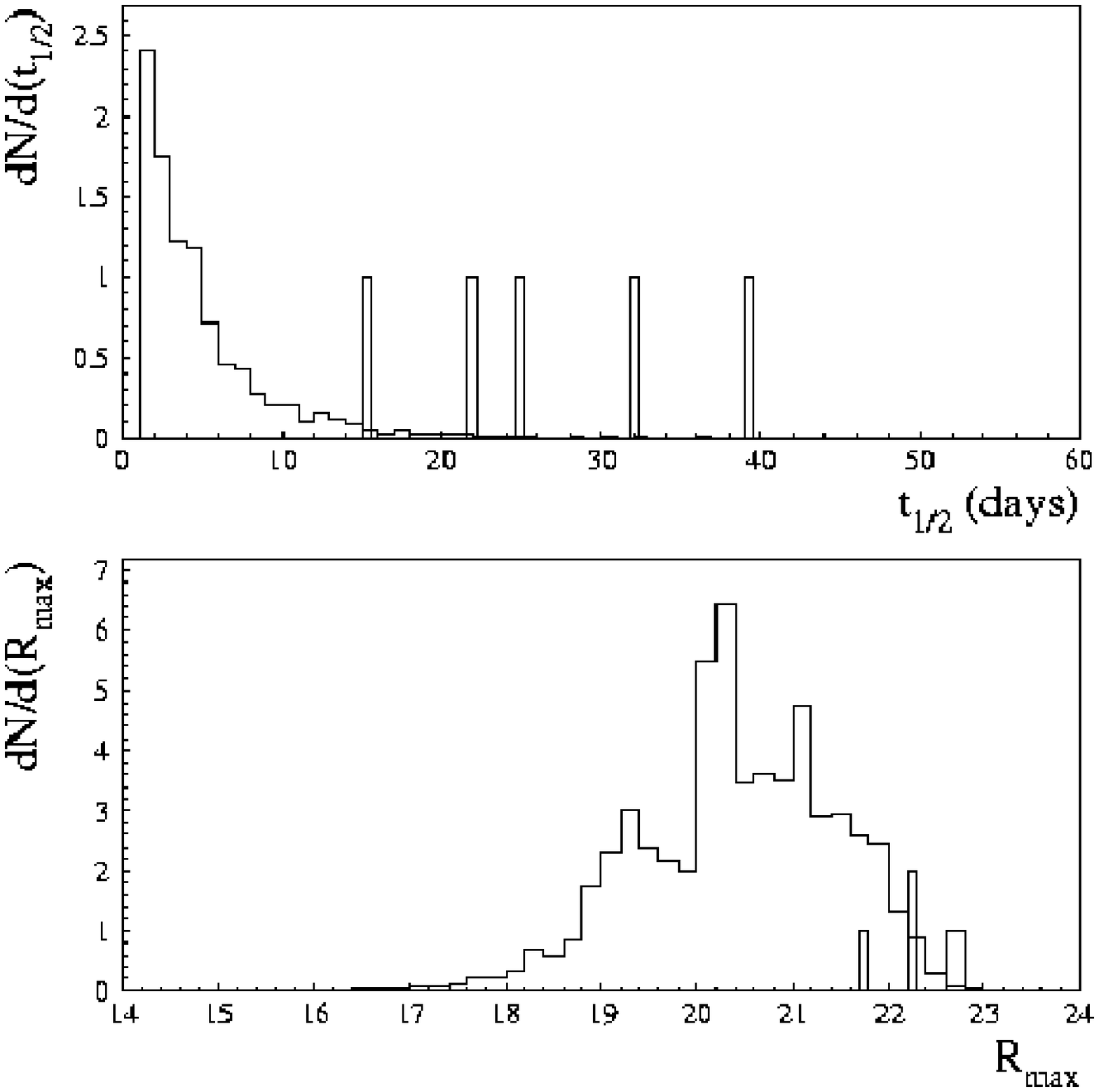}}
\parbox[b]{7 cm} \vfill
{\caption{Results of Monte Carlo simulations,
$m_{MACHO}^{halo}=10^{-2}\,M_\odot$ case. The scale on the y
coordinate of the two distributions on the right are in arbitrary
units. The dots (left) give the position of the candidates
as labelled in Table \ref{tavola2}.}  \label{fig:mc001}}
\end{figure*}

\begin{figure*}
\resizebox{7 cm}{!}
{\includegraphics{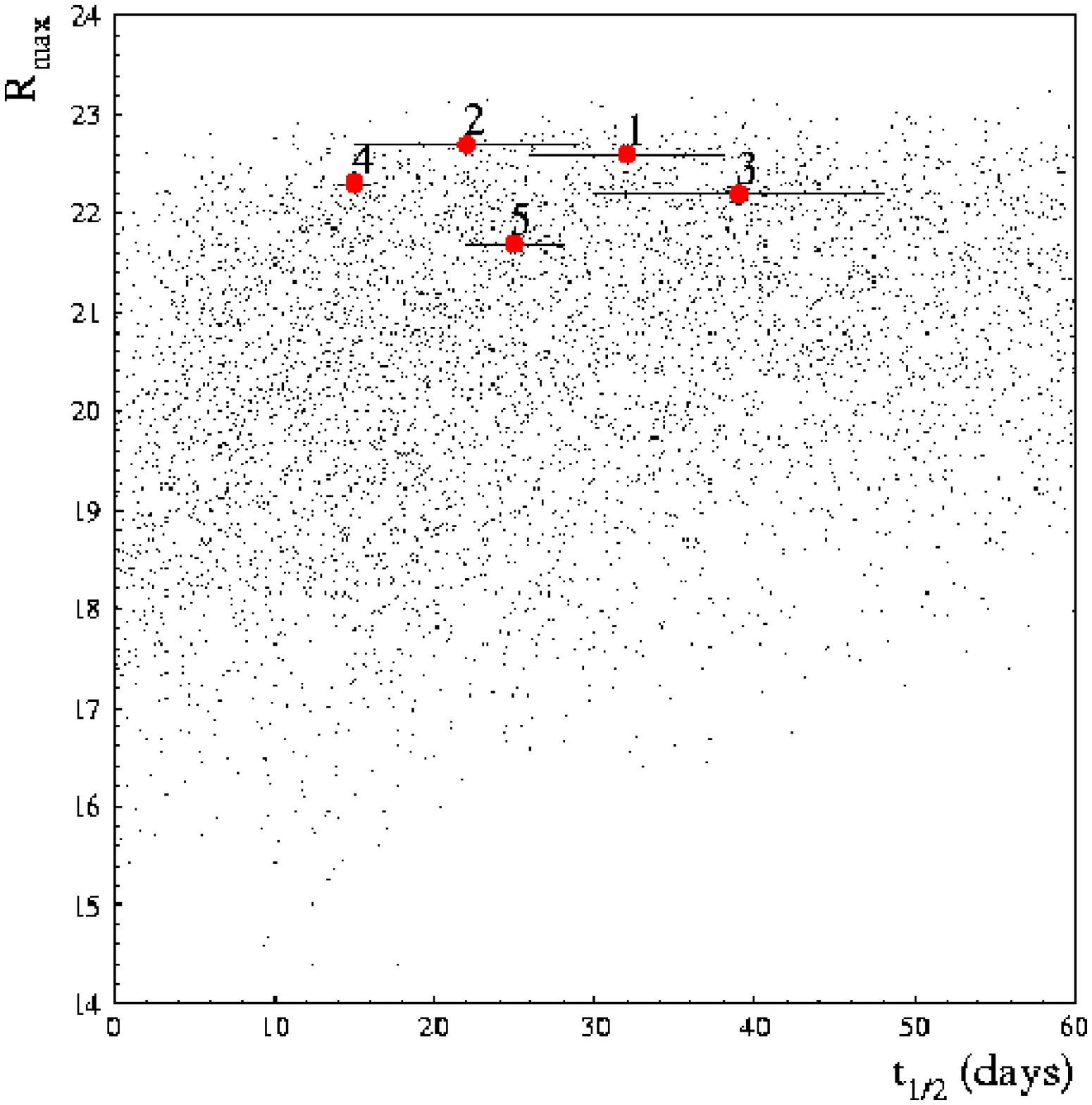}}
\resizebox{7 cm}{!}
{\includegraphics{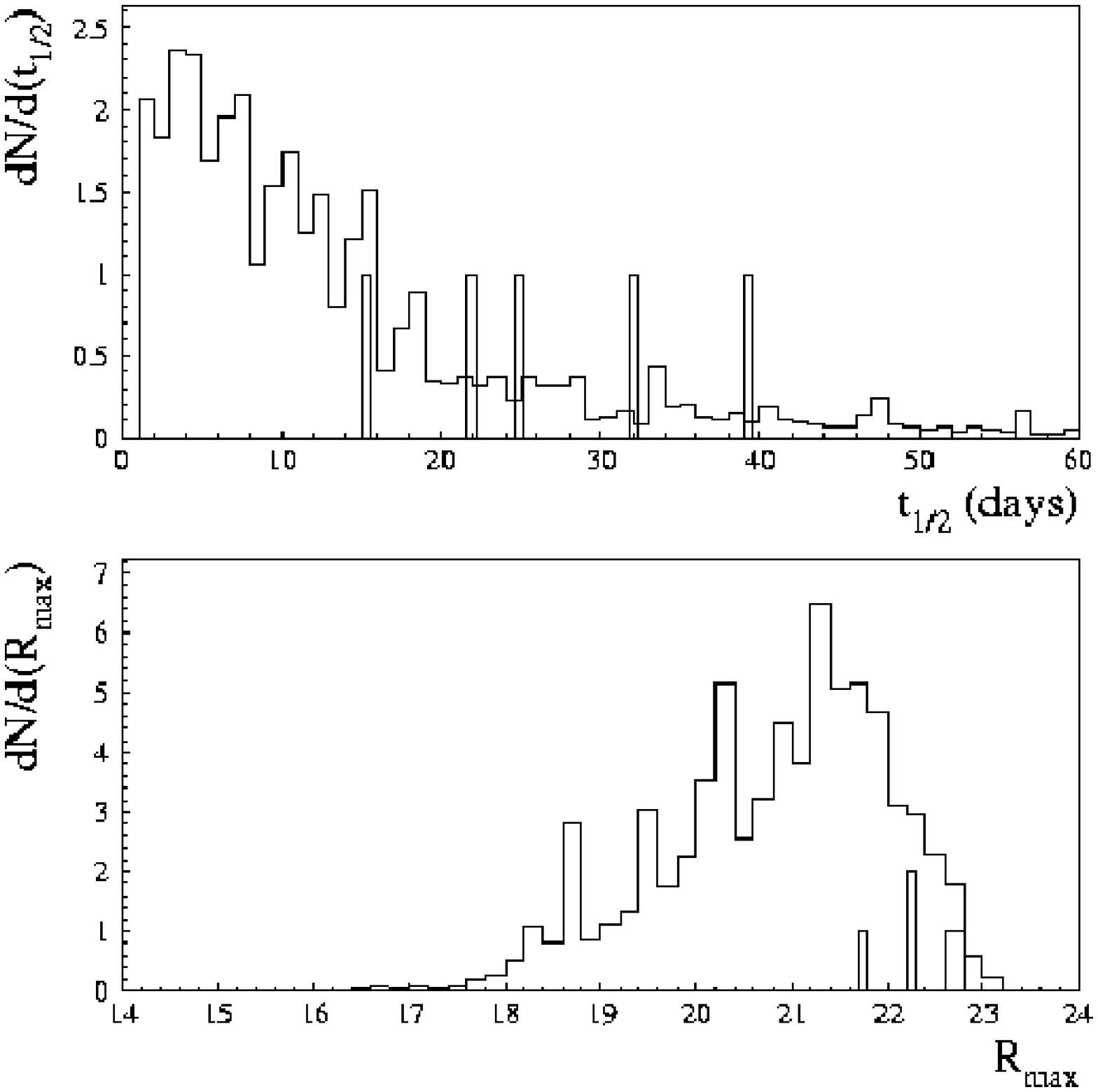}}
\parbox[b]{7 cm} \vfill
{\caption{Results of Monte Carlo simulations,
$m_{MACHO}^{halo}=0.5\,M_\odot$ case. The scale on the y
coordinate of the two distributions on the right are in arbitrary
units. The dots (left) give the position of the candidates
as labelled in Table \ref{tavola2}.} \label{fig:mc05}}
\end{figure*}

Locating our candidates in the parameter space predicted by the
simulation is more meaningful. We report the expected values of
$t_{1/2}$ and on the $R$ magnitude at maximum. In Figs.
\ref{fig:mc001} and \ref{fig:mc05} we give plots of their
functional relation and of their projected distributions. On
these same plots we give the position of our selected light
curves in this parameter space.

Looking at the distributions we  notice that for the
$m_{MACHO}^{halo}=0.5\,M_\odot$ and
$m_{MACHO}^{halo}=0.01\,M_\odot$ cases, $80\%$ of the light curves
are expected to have a time width at half maximum $t_{1/2}<24$ and
$t_{1/2}<10$ days, respectively (of course, shorter events are
expected when the MACHO  mass is smaller). In both cases, $\sim
80\%$ of the events are predicted with a magnitude at maximum
$R_{max}<21.7$. Our candidates have $t_{1/2}\ge 15$ days and
$R_{max}\ge 21.7$ (somewhat at the limit of the expected
distributions) and therefore most of them are probably not
microlensing events. Still, we expect $\sim 1$ self--lensing
event and it is possible that one or two of them are true
microlensing events. In any case, from the $t_{1/2}$
distribution, we are led to exclude that the microlensing
candidate events are due to MACHOs of very small mass (only $\sim
10\%$ of events with $m_{MACHO}^{halo}=0.01\,M_\odot$ are expected
to have $t_{1/2}>15$ days). This is indeed in agreement with the
results found by the MACHO and EROS collaborations: they  find
lens masses in the halo within the range  0.2--0.6 $M_\odot$
(Alcock et al. 2000; Lasserre et al. 2000). 

We are not yet in a position to tell what kind
of varying objects generate our events if they are not
due to microlensing. They may be irregular or long
period variable giants, but only a longer time baseline,
and/or observations of the object at minimum light,
will allow us to conclude.

The MDM analysis is not yet complete. The results from the
analysis of data acquired in the other field (located on the
opposite side with respect to the major axis of M31 of the Target
field analysed here) and results from new  observations scheduled
for the fall 2001 will give us the opportunity to gain further
insight into the still open question of the composition of dark
haloes. At the present time, with the caution suggested by the
just mentioned problems, the analysis discussed in this paper
tends to confirm that only a minor fraction of dark matter in the
galactic haloes is in form of MACHOs  within the mass range 0.01--0.5 
$M_\odot$ under the assumption of a standard model of the halo
and given source luminosity functions.

\begin{acknowledgements}
We thank M. Cr\'{e}z\'{e}, S. Droz, L. Grenacher, G. Marmo, G.
Papini and N. Straumann for useful discussions and suggestions.
Work by AG was supported by NSF grant AST $\sim$97-27520 and by a
grant from Le Centre Fran\c{c}ais pour l'Accueil et les Echanges
Internationaux.
\end{acknowledgements}

\end{document}